\documentstyle{mn} 
\input epsf
\title[Galaxies in Abell~2634]
{X-ray emission from the galaxies in Abell~2634}
\author[I.~Sakelliou and M. R. Merrifield]
          {I.~Sakelliou, M. R.~Merrifield \\
          Department of Physics and Astronomy, University of Southampton,
          Southampton SO17 1BJ}
\begin{document}
\maketitle

\begin{abstract} 

It is difficult to detect X-ray emission associated with galaxies in
rich clusters because the X-ray images of the clusters are dominated
by the emission from their hot intracluster media (ICM).  Only the
nearby Virgo cluster provides us with information about the X-ray
properties of galaxies in clusters.  Here we report on the analysis of
a deep {\it ROSAT} HRI image of the moderately-rich cluster, Abell
2634, by which we have been able to detect the X-ray emission from
the galaxies in the cluster.  The ICM of Abell~2634 is an order
of magnitude denser than that of the Virgo cluster, and so this
analysis allows us to explore the X-ray properties of individual
galaxies in the richest environment yet explored.

By stacking the X-ray images of the galaxies together, we have shown
that their emission appears to be marginally resolved by the HRI.
This extent is smaller than for galaxies in poorer environments, and
is comparable to the size of the galaxies in optical light.  These
facts suggest that the detected X-ray emission originates from the
stellar populations of the galaxies, rather than from extended hot
interstellar media.  Support for this hypothesis comes from placing
the optical and X-ray luminosities of these galaxies in the $L_{\rm
B}$--$L_{\rm X}$ plane: the galaxies of Abell~2634 lie in the region
of this plane where models indicate that all the X-ray emission can be
explained by the usual population of X-ray binaries.  It is therefore
probable that ram pressure stripping has removed the hot gas component
from these galaxies.

\end{abstract}


\section{Introduction}

The advent of the {\it Einstein} observatory changed the belief that
early-type galaxies contain little interstellar gas by revealing hot
X-ray emitting halos associated with many of them (e.g.  Forman et
al. 1979).  Subsequent X-ray observations led to the conclusion that
these galaxies can retain large amounts (up to $\sim 10 ^{11}
M_{\odot}$) of hot ($T \sim 10^{7}$ K) interstellar medium
(Forman, Jones \& Tucker 1985; Trinchieri \& Fabbiano 1985; Canizares,
Fabbiano \& Trinchieri 1987).

\begin{figure*}
\begin{center}
 \leavevmode
 \epsfxsize 0.6\hsize
 \epsffile{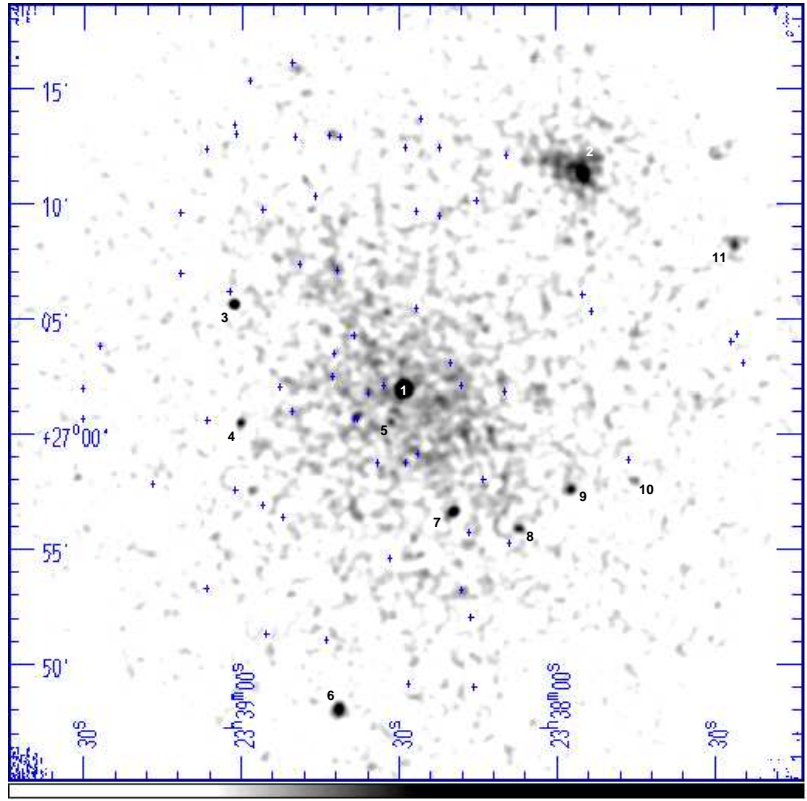}
 \caption{Grey-scale {\it ROSAT} HRI image of the core of the cluster
Abell~2634. The image has been smoothed using a Gaussian kernel with a
dispersion of 8 arcsec. The positions of the cluster galaxies with
measured redshift are marked with crosses.} 
\end{center}
\label{grayscale}
\end{figure*}

However, this picture might be expected to be different for galaxies
that reside near the centres of rich clusters of galaxies, since their
properties must be affected by their dense environment.  For example,
we might expect interstellar medium (ISM) gas to be stripped from the
galaxy by the ram pressure resulting from the passage of the galaxy
through the intracluster medium (ICM) (Gunn \& Gott 1972).  Stripping
of the ISM can also result from tidal interactions with other nearby
galaxies (Richstone 1975; Merritt 1983, 1984).  The most dramatic and
well-studied example of a galaxy which appears to be in the process of
being stripped of its ISM is the elliptical galaxy M86 in the Virgo
cluster, which shows a `plume' of X-ray emission emanating from it
(Forman et al. 1979; White et al. 1991; Rangarajan et al. 1995).

In addition to the mechanisms which remove the ISM of a galaxy, gas
can also be replenished.  The gravitational pull of a galaxy attracts
the surrounding ICM.  This gas ends up being concentrated in or behind
the galaxy, depending on the velocity of the galaxy relative to the
ICM (see, for example, Sakelliou, Merrifield \& McHardy 1996). Stellar
winds can also replenish the hot gas in a galaxy's ISM.

All the processes mentioned above take place simultaneously. The
relative importance of each process depends on: the galaxies'
velocities; the local density of the ICM; the number density of
galaxies; their orbits in the cluster; and the gravitational potential
of each galaxy.  It is therefore {\it a priori} difficult to say which
mechanism dominates in the cores of rich clusters of galaxies, and
hence whether cluster galaxies are surrounded by the extensive X-ray
emitting halos that we see associated with galaxies in the field.

Unfortunately, X-ray observations of rich clusters have generally not
been of high enough quality to answer this question, since any
emission from the galaxies is hard to detect against the high X-ray
background produced by the cluster's ICM (Canizares \& Blizzard 1991;
Vikhlinin, Forman \& Jones 1994; Bechtold et al. 1983; Grebenev et
al. 1995; Soltan \& Fabricant 1990; Mahdavi et al. 1996). In the cases
where galaxy X-ray emission has been reported, the studies have been
restricted to a few bright cluster galaxies, and it has not proved
possible to investigate the general galaxy population in a
statistically complete manner.

In order to search for X-ray emission from galaxies in a moderately
rich environment, we obtained a deep {\it ROSAT} HRI observation of
the core of the rich cluster Abell~2634, which is a nearby (z=0.0312)
centrally-concentrated cluster of richness class I.  In \S2.1 we
describe the analysis by which the X-ray emission from the galaxies in
this cluster was detected.  In \S2.2 we explore the properties of this
X-ray emission, and show that the galaxies in this cluster lack the
extensive gaseous halos of similar galaxies in poorer environments.
In \S3, we show how this difference can be attributed to ram
pressure stripping.

\section{X-ray observations and analysis}

The core of Abell~2634 was observed with the {\it ROSAT} HRI in two
pointings, in January and June 1995, for a total of $62.5\,{\rm
ksec}$.  The analysis of these data was performed with the IRAF/PROS
software.

Inspection of the emission from the cD galaxy and other bright X-ray
sources in the images from the two separate observations indicates
that the two sets of observations do not register exactly and that a
correction to the nominal {\it ROSAT} pointing position is
required. Therefore, the second set of observations was shifted by
$\sim$ 2.0 arcsec to the east and $\sim$ 0.8 arcsec to the south; such
a displacement is consistent with typical {\it ROSAT} pointing
uncertainties (Briel et al. 1996).  Both images were then registered
with the optical reference frame to better than an arcsecond.  A
grey-scale image of the total exposure is shown in Fig. 1.  The image
has been smoothed with a Gaussian kernel of 8 arcseconds dispersion.
At the distance of Abell~2634, $1\,{\rm arcsec}$ corresponds to
$900\,{\rm pc}$.\footnote{Here, as throughout this paper, we have
adopted a Hubble constant of $H_{0}=50\; {\rm km\:s^{-1}\:Mpc^{-1}}$.}

\begin{table}
 \caption{Bright Sources}
 \begin{tabular}{cccc}
\hline \hline
Source & $\alpha$(J2000) & $\delta$(J2000) & ID/notes \\
     & $^{\rm h} \; ^{\rm m} \; ^{\rm s} $ & $\degr \; \arcmin \;
\arcsec$ & \\
\hline
1 & 23 38 29.1 & 27 01 53.5 & cD galaxy \\
2 & 23 37 56.1 & 27 11 31.3 & cluster \\
3 & 23 39 01.6 & 27 05 35.9 & star \\
4 & 23 39 00.5 & 27 00 27.9 & star\\
5 & 23 38 31.7 & 27 00 30.5 & nothing \\
6 & 23 38 41.5 & 26 48 04.1 & star \\
7 & 23 38 19.8 & 26 56 41.5 & ? \\
8 & 23 38 07.4 & 26 55 52.8 & star \\
9 & 23 37 57.5 & 26 57 30.1 & galaxy ?\\
10 & 23 37 45.3 & 26 57 53.1 & two objects \\
11 & 23 37 26.2 & 27 08 14.6 & ? \\

\hline
 \end{tabular}
\end{table}

This deep image of Abell~2634 reveals the largescale X-ray emission
from the hot ICM of the cluster and a few bright sources, which are
numbered on Fig.~1.  Source 1 is the cD galaxy NGC~7720, located near
the centre of Abell~2634. It hosts the prototype wide-angle tailed
radio source 3C~465 (e.g.~Eilek et al.  1984).  Source 2 is a
background cluster at a redshift of $cz \simeq 37,000 \ {\rm km \
s^{-1}}$ (Pinkney et al.  1993; Scodeggio et al.  1995).  For the rest
of the X-ray bright sources, the Automatic Plate Measuring 
machine, run by the Royal Greenwich Observatory in Cambridge, was used
to obtain optical identifications.  Table 1. gives the positions of
these sources as determined from the X-ray image, and the class of
their optical counterparts.  The position of source 7 coincides with a
faint object in the Palomar sky survey, but there is also a nearby
star, and source 11 does not seem to have a discernible optical
counterpart.  All these sources were masked out in the subsequent
analysis.  

The positions of galaxies that are members of Abell~2634 are also
indicated on Fig.~1.  Pinkney et al.\ (1993) collected the redshifts
of $\sim$150 galaxies that are probable members Abell~2634 (on the
basis that their redshifts lie in the range $6,000 < cz < 14,000\,
{\rm km\,s^{-1}}$), and Scodeggio et al. (1995) have increased the
number of galaxies whose redshifts confirm that they are cluster
members up to $\sim 200$.  The sample of redshifts is complete to a
magnitude limit of 16.5, and from this magnitude-limited sample we
have selected those galaxies that appear projected within a circle of
15 arcmin radius, centered on the cD galaxy.  This selection yields 62
galaxies, of which the vast majority are of type E and S0 -- only 10
are classified as spirals or irregular.  The positions of these
galaxies are taken from the CCD photometry of Pinkney (1995) and
Scodeggio et al. (1995), and are accurate to $\sim 1\,{\rm arcsec}$.
They are marked as crosses on Fig.~1.

Inspection of Fig.~1 reveals several cases where the location of a
galaxy seems to coincide with an enhancement in the cluster's X-ray
emission, and it is tempting to interpret such enhancements as the
emission from the galaxy's ISM.  However, it is also clear from Fig.~1
that the X-ray emission in this cluster contains significant
small-scale fluctuations and non-uniformities.  We must therefore
consider the possibility that the apparent associations between galaxy
locations and local excesses in the X-ray emission may be
chance superpositions.  We therefore now present a more objective
approach to searching for the X-ray emission from cluster galaxies.

\subsection{Detection of the cluster galaxies}

Before adopting an approach to detecting the emission from cluster
galaxies, we must first have some notion as to how bright we might
expect the emission to appear in this deep HRI image.  Previous X-ray
observations have shown that the X-ray luminosities of E and S0
galaxies in the 0.2-3.5 keV energy band range from $\sim10^{39}$ to
$\sim 10^{42} \ {\rm erg \ s^{-1}}$ (Kim, Fabbiano \& Trinchieri
1992a, b; Forman et al. 1985).  These limits at the distance of
Abell~2634 correspond to fluxes of $5 \times 10^{-16}$ to $5 \times
10^{-13} \ {\rm erg \ s^{-1} \ cm^{-2}}$.  We have used the PIMMS
software to convert these limits to count rates for the {\it ROSAT}
HRI detector.  The emission from the galaxies was modeled by a
Raymond-Smith plasma (Raymond \& Smith 1977) with a temperature
$kT=0.862$ keV and a metal abundance of 25\% solar; these quantities
are consistent with the values previously found from observations of
early-type galaxies (Kim et al.  1992a; Matsushita et al. 1994; Awaki
et al. 1994). The absorption by the galactic hydrogen was also taken
into account by using the column density given by Stark et al. (1992)
for the direction of Abell~2634 ($N_{\rm H}= 4.94 \times 10^{20} \
{\rm cm^{-2}}$).  These calculations predict that the $62.5\,{\rm
ksec}$ HRI observation of this cluster should yield somewhere between
$\sim1$ and $\sim1200$ counts from each galaxy.  Motivated by this
prediction of a respectable, but not huge, number of counts per
galaxy, we set out to detect emission associated with cluster
galaxies.

We are trying to detect this fairly modest amount of emission against
the bright background of the ICM emission.  We therefore seek to
improve the statistics by stacking together the X-ray images in the
vicinity of the 40 E and S0 galaxies marked in Fig.~1.
Fig. 2. presents a contour plot of the combined image, which covers a
region of 1 arcmin radius around the stacked galaxies.  The centre of
the plot coincides with the optical centres of the individual
galaxies. Clearly, there appears to be X-ray emission associated with
the cluster galaxies, and it is centered at their optical positions.
This coincidence provides us with some confidence that the X-ray and
optical frames are correctly registered.  We have also constructed a
composite brightness profile for the 40 galaxies by adding the
unsmoothed counts detected in concentric annuli centered on each
galaxy.  The width of each annulus in this profile was set at 6 arcsec
and the local background, as measured in an annulus between 1.0 and
2.0 arcmin around each galaxy, was subtracted.  The resulting profile
is presented in Fig. 3.  Once again, the excess of emission in the
vicinity of the cluster galaxies is apparent.

In order to assess the significance of this detection, we generated
100 sets of simulated data from randomly selected points on the image.
The diffuse emission from the ICM varies systematically with radius,
and so we might expect the probability that a galaxy is coincidentally
aligned with a clump in the ICM emission to vary systematically with
radius.  Further, the sensitivity of the HRI varies with radius, and
so the detectability of the emission from a single galaxy will vary
with radius as well.  We therefore constructed the simulated data sets
by extracting counts from the HRI image at the same radii as the true
galaxy locations, but at randomized azimuthal angles.  The mean
profile and the RMS fluctuations amongst the simulated data sets are
shown in Fig.~3.  As might be expected, the average number of counts
in these random data sets is zero; the larger RMS error bars at small
radii reflect the smaller sizes of these annuli.  From a $\chi^{2}$
comparison between the observed galaxy profile and the simulated
profile, we can conclude that there is less than 0.1\% probability
that the apparent peak in the galaxy emission is produced by chance.
Thus, the detection of emission from the galaxies in Abell~2634 is
highly statistically significant.

\begin{figure}
\begin{center}
 \leavevmode
 \epsfxsize 0.9\hsize
 \epsffile{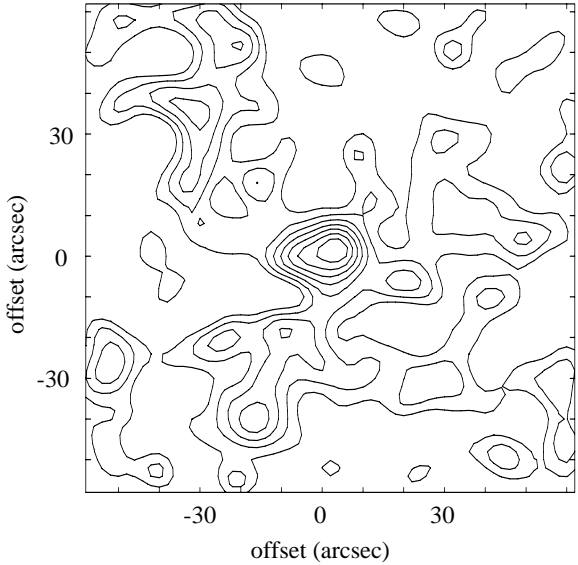}
 \caption{Contour plot of the combined image of all the early-type
galaxies that belong to Abell~2634. The pixel size of the image is 2
arcsec and it has been smoothed with a Gaussian kernel of 2
pixels. The center of the plot coincides with the optical centres of
the galaxies. The contour lines are from 20 to 100 per cent the peak
value and are spaced linearly in intervals of 5 per cent.}
\end{center}
\label{contour}
\end{figure}

\begin{figure}
\begin{center}
 \leavevmode
 \epsfxsize 0.9\hsize
 \epsffile{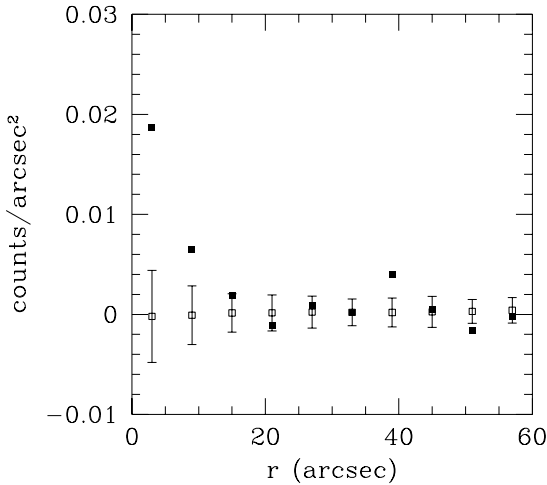}
 \caption{The combined surface brightness profile of all the
early-type galaxies (filled squares) normalized to one galaxy. Open squares
represent the average profile from  the simulations.} 
\end{center}
\label{earlysimul}
\end{figure}

\subsection{Origin of the X-ray emission}

As mentioned in the introduction, early-type galaxies have been found
to retain large amounts of hot gas, which extends far beyond the
optical limits of the galaxies. X-ray binaries also contribute to the
total emission, and they become more dominant in X-ray faint galaxies.
We might also expect some of the emission to originate from faint
active galactic nuclei (AGNs) in the cores of these galaxies.
Although none of the galaxies in our sample has been reported as an
active galaxy, there is increasing dynamical evidence that the vast
majority of elliptical galaxies contain central massive black holes
(van der Marel et al. 1997; Kormendy et al.  1996a, 1996b; for a
review Kormendy \& Richstone 1995), and so we might expect some
contribution from low-level activity in such systems.  We therefore
now see what constraints the observed X-ray properties of the galaxies
in Abell~2634 can place on the origins of the emission.

\subsubsection{The extent of the X-ray emission}

One diagnostic of the origins of the X-ray emission is the measurement
of its spatial extent.  AGN emission should be unresolved by the HRI,
while emission from X-ray binaries should be spread over a similar
spatial scale as the optical emission, and halos of hot gas should be
still more extended.

\begin{figure*}
\begin{center}
 \leavevmode
 \epsfxsize 0.45\hsize
 \epsffile{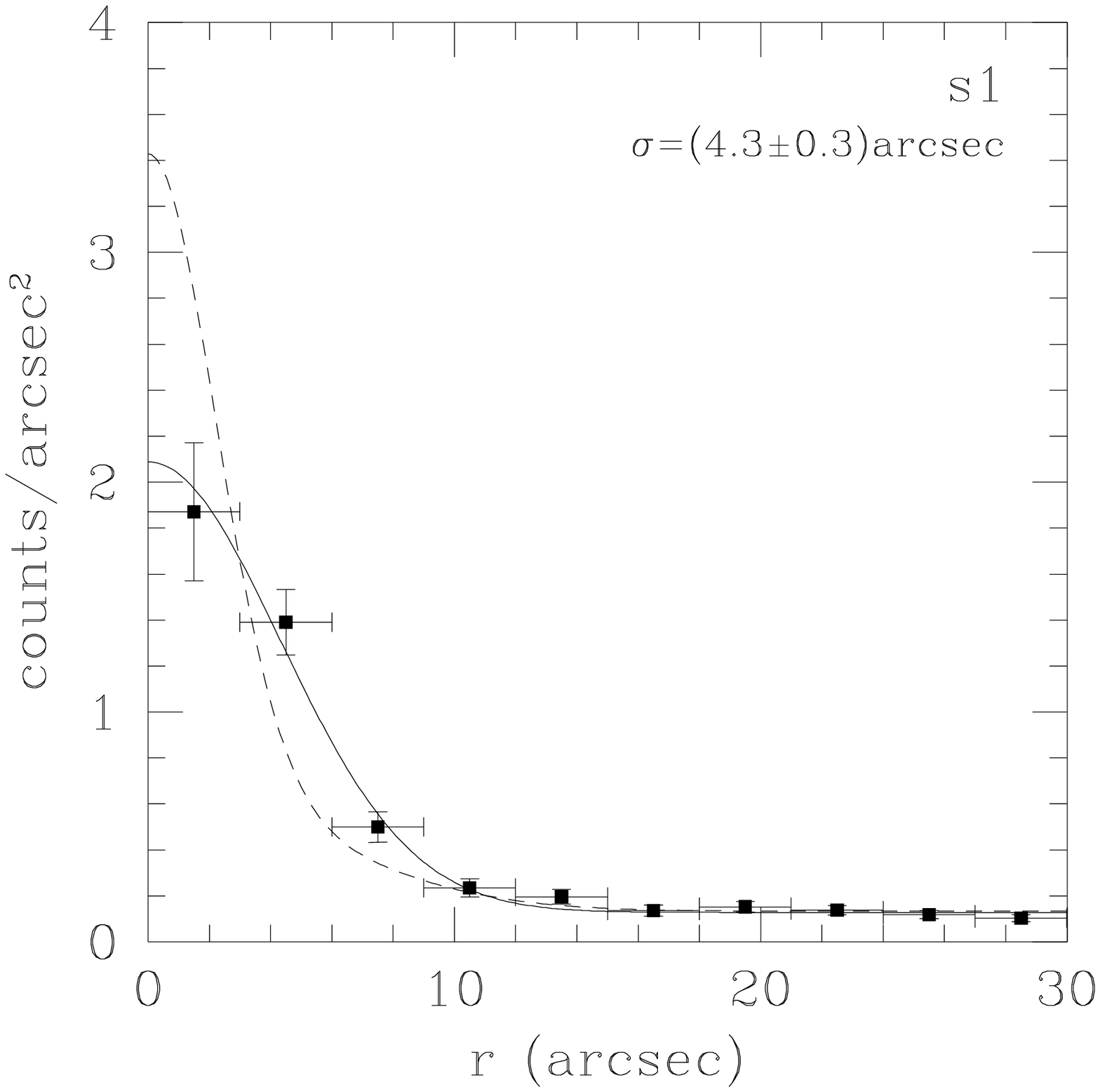}
 \epsfxsize 0.45\hsize
 \epsffile{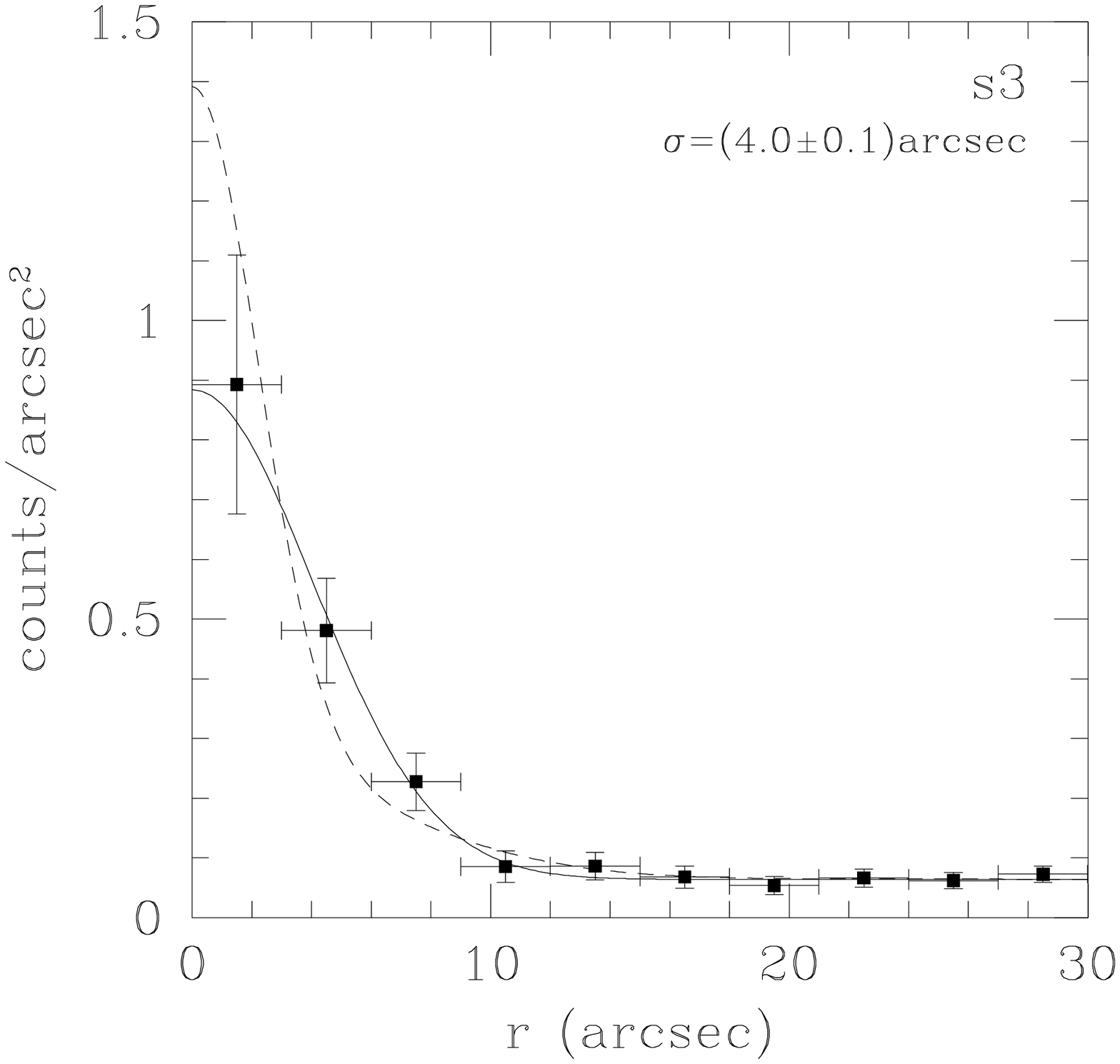}
\end{center}
\begin{center}
 \leavevmode 
 \epsfxsize 0.45\hsize
 \epsffile{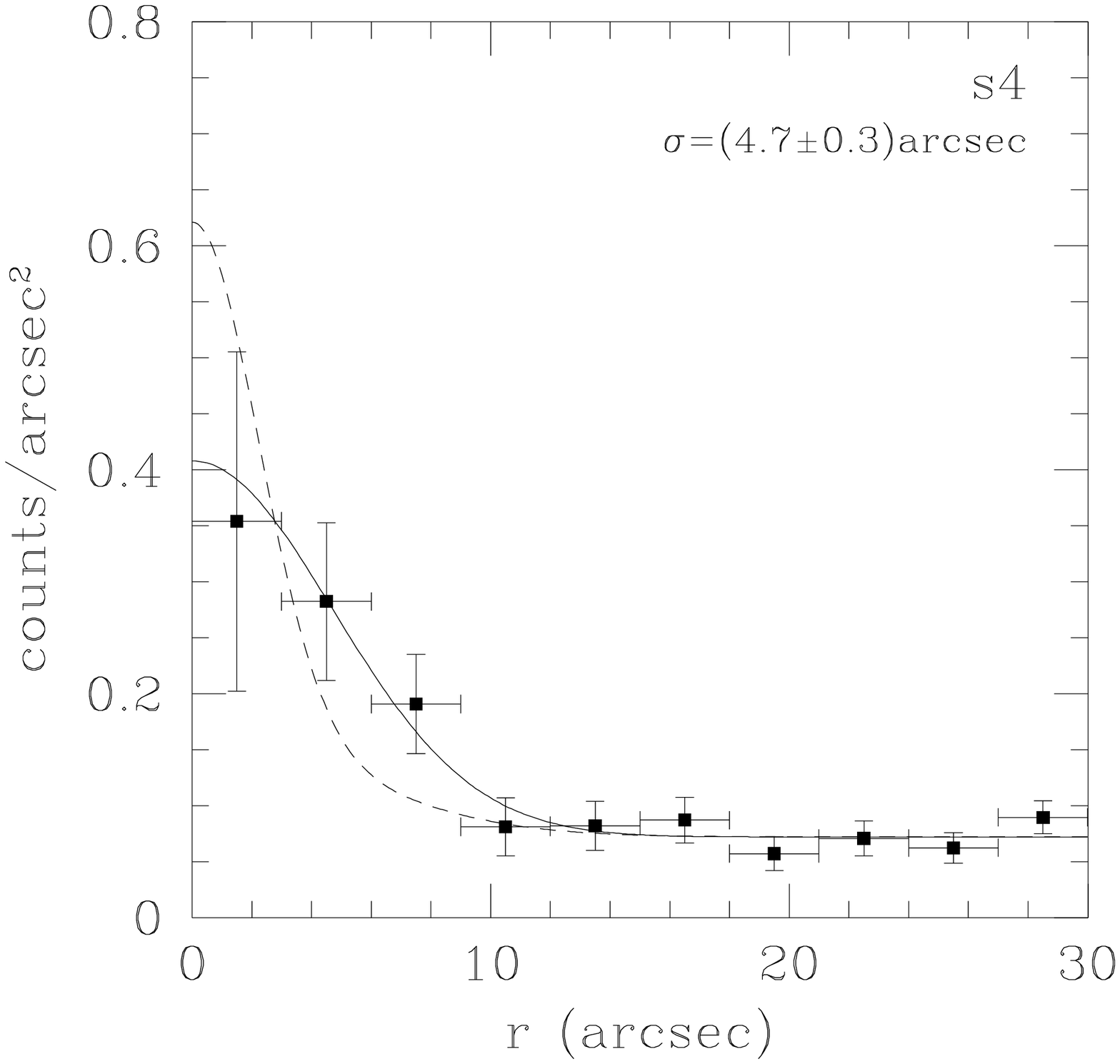}
 \epsfxsize 0.45\hsize
 \epsffile{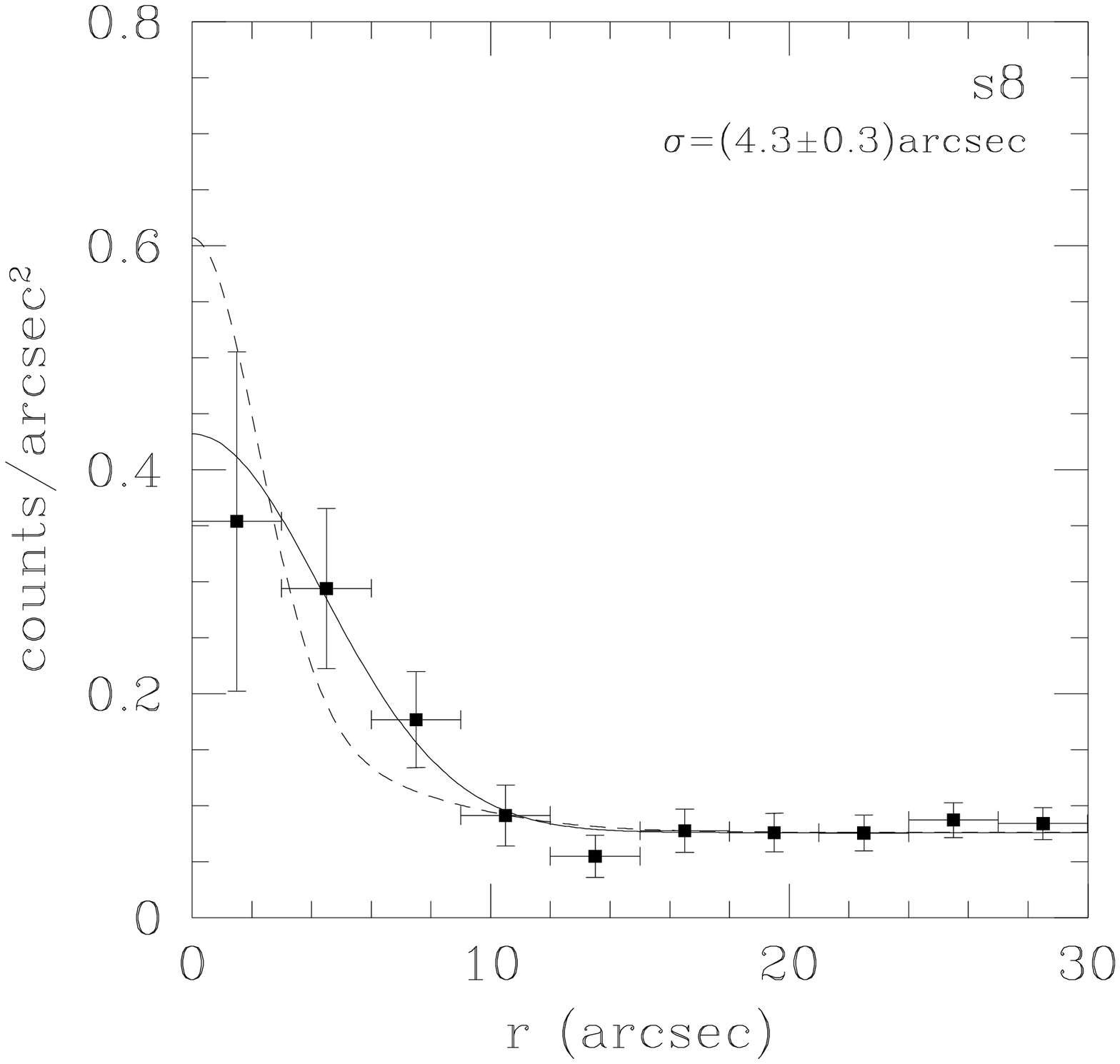}
\end{center}
\caption{Surface brightness distribution of the bright sources in the
HRI image. (s1, s3, s4, and s8). Source 1 is the central cD
galaxy. The profile is fitted by the appropriate HRI PSF for the
distance of the point source from the centre of the image (dash line)
and a Gaussian (solid line). The calculated width ($\sigma$) of the
best fit Gaussian is also given.}
\label{sources}
\end{figure*}
 
In order to assess the spatial extent of the X-ray emission, we need
to characterize the PSF in this HRI observation.  The point sources
detected in these data are more extended than the model PSF for the
HRI detector given by Briel et al. (1996) as is seen in Fig.~4, where
four of the sources are fitted by this model PSF (dash line).  This
discrepancy can be attributed to residual errors in the reconstruction
of {\it ROSAT}'s attitude, which broaden the PSF in long integrations.
We have therefore empirically determined the PSF that is appropriate
for this observation by fitting the profiles of the point sources with
a Gaussian PSF model (Fig.~4, solid line). Only sources 1, 3, 4, 8 are
used for the determination of the width of the Gaussian.  Source 6 is
very elongated and can not be represented by a symmetrical
function. The mean dispersion of the best-fit model was found to be $(4.1
\pm 0.1)$ arcsec.  All of the point sources detected in this image
have widths consistent with this value, and so there is no evidence
that the PSF varies with radius.  We therefore adopt this PSF for the
emission from all the galaxies in the observation.

Figure~5 shows the comparison between the adopted PSF and the emission
from the cluster galaxies.  The emission appears to be more extended
than the PSF; fitting the data to the PSF yields a $\chi^{2}$ value of
14.2 with 9 degrees of freedom, which is marginally consistent with
the emission being unresolved.  We can obtain a better fit by modeling
the radial profile of the emission using a Gaussian, which we convolve
with the PSF to model the observed profile.  Fitting this model to the
observations, we find that the intrinsic width of the X-ray emission
is $4.3^{+2.2}_{-2.8}$ arcsec. The best-fit model is also shown in
Fig.~5.

\begin{figure}
\begin{center}
 \leavevmode
 \epsfxsize 0.9\hsize
 \epsffile{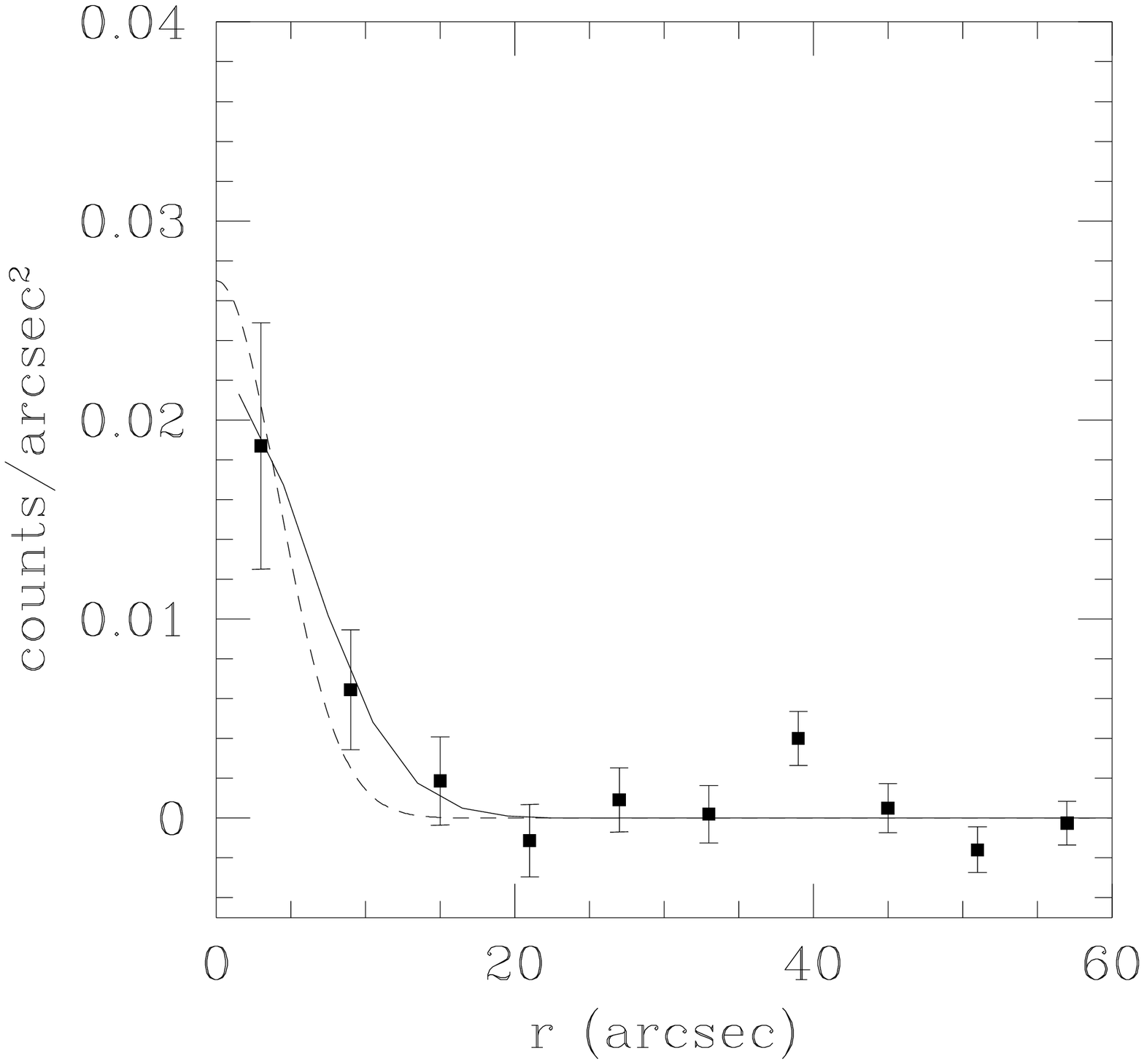}
\caption{The combined surface brightness distribution of the
40 early-type galaxies, normalized to one galaxy. The profile is fitted
by the measured HRI PSF (dashed line) and the spatially-extended model 
(solid line).}
\end{center}
\label{earlyprof}
\end{figure}

The radius of the X-ray halos of early-type galaxies with optical
luminosities comparable to those in this cluster has been shown to be
$\sim 20$ -- 60 kpc (e.g.~Fabbiano et al. 1992), with the lower values
characterizing optically fainter galaxies. At the distance of
Abell~2634 these values correspond to $\sim 20$ -- 60 arcsec, much
larger than the upper limit of $\sim$ 7 arcsec we found for the extent
of the galactic X-ray emission.  Thus, the X-ray emission from the
cluster galaxies, although apparently extended, clearly does not
originate from the large halos of hot gas found around comparable
galaxies in poorer environments.

One possible explanation for the spatial extent of the emission from
these galaxies is that it could arise from errors in the adopted
positions for the galaxies.  Such errors would broaden the
distribution of X-rays when the data from different galaxies are
co-added even if the individual sources are unresolved.  However, the
zero-point of the X-ray reference frame is well tied-down by
the detected point sources in the field.  Further, the optical
locations of the galaxies come from CCD photometry with positional
errors of less than an arcsecond.  It therefore cannot explain the
$\sim 4$ arcsec extent of the observed X-ray emission.

We therefore now turn to the extent of the X-ray emission that we
might expect from X-ray binaries.  Nearly half of the early-type
galaxies that we use for our analysis have been imaged in the I-band
by Scodeggio, Giovanelli \& Haynes (1997).  They have fitted the
optical galaxy profile with a de Vaucouleur law, and found a mean
value for their effective radii of $\sim$8 arcsec, with only 4
galaxies smaller than 3 arcsec and another 3 larger than 13 arcsec.
These values are directly comparable to the spatial extent of the
X-ray emission derived above.  Thus, it would appear that the
observations are consistent with what we would expect if the X-ray
emission from the galaxies in Abell~2634 originates from X-ray
binaries in these systems, although we have not ruled out the
possibility that some fraction of the emission comes from AGN.

\subsubsection{The luminosity of the X-ray emission}

A further test of the origins of the X-ray emission in the cluster
galaxies comes from its luminosity.  It has been found that the blue
luminosities of galaxies correlates with their X-ray luminosities,
with the optically brighter galaxies being more luminous in X-rays
(e.g.~Forman et al. 1985; Fabbiano et al.  1992).  This correlation
for the early-type galaxies in the Virgo cluster is presented in
Fig. 6. The optical and X-ray luminosities of these galaxies are taken
from Fabbiano et al. (1992).  The line in this plot divides the
$L_{\rm B} - L_{\rm X}$ plane into two distinct galaxy types (Fabbiano
\& Schweizer 1995).  In addition to the differences in the ratio of
X-ray-to-optical luminosities, galaxies in these two regions have been
shown to possess different spectral properties.  The spectra of the
X-ray bright galaxies [group (I)] are well fitted by Raymond-Smith
models of 1 keV temperature, and it is believed that these galaxies
retain large amounts of hot ISM.  In the spectra of the X-ray faint
galaxies of group (II), on the other hand, a hard component is
present; X-ray binaries are believed to be the major source of the
X-rays in these galaxies.

\begin{figure}
\begin{center}
 \leavevmode
 \epsfxsize 0.9\hsize
 \epsffile{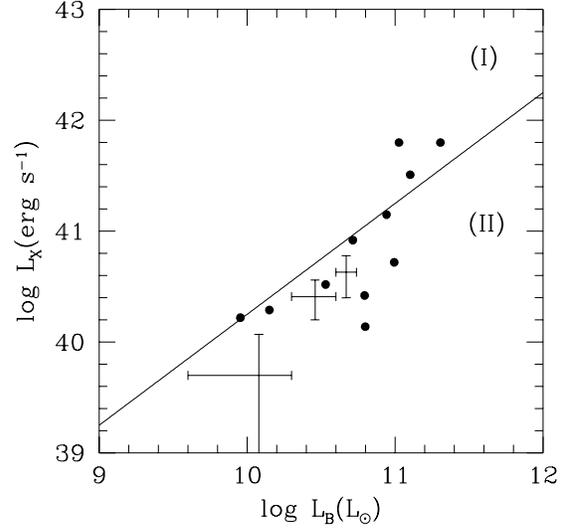}
\caption{X-ray luminosity versus blue luminosity for the early-type
galaxies. The line delineates the boundary between the locations of
galaxies where the hot ISM makes a significant contribution to the
total emission [region (I)], and the locations of galaxies where the
entire emission can be ascribed to X-ray binaries [region (II)].  The
locations in this plane of galaxies that belong to the Virgo cluster
are marked by filled circles.  The average properties of galaxies in
Abell~2634 lying in
different optical luminosity ranges are indicated by crosses.}
\end{center} 
\label{LxLB}
\end{figure}

In order to see where the galaxies of Abell~2634 lie in this plot, we
must calculate their optical and X-ray luminosities.  Butcher \&
Oemler (1985) have measured {\it J} and {\it F} optical magnitudes for
a large number of galaxies in Abell~2634.  We have converted these
magnitudes to the blue band by applying the colour relations provided
by Oemler (1974) and Butcher \& Oemler (1985), correcting for galactic
extinction, and using the appropriate K-correction.  We find that the
absolute blue magnitude of the galaxies in the HRI image lie in the
range from $-18.5$ to $-21.4$.  We have divided these galaxies into
three groups according to their optical luminosities: group A
$(0.4-2.0) \times 10^{10} \ {\rm L_{\sun}}$ with 17 galaxies; group B
$(2.1-3.7) \times 10^{10} \ {\rm L_{\sun}}$ with 8 galaxies; and group
C $(3.8-5.5) \times 10^{10} \ {\rm L_{\sun}}$ with only 2 galaxies.

The X-ray luminosity of each group was obtained by repeating the
analysis of \S2.1 using just the galaxies in each sub-sample.  Using
the PIMMS software we converted the observed count rate from the HRI
image to X-ray luminosity in the energy range 0.2-3.5 keV.  The thermal
model used for the conversion is the same that was used by Fabbiano et
al. (1992) to derive the plot shown in Fig. 6, and is discussed in
\S2.1.   

The resulting values for optical and X-ray luminosities in each
sub-sample are shown in Fig.~6.  The horizontal error bars represent
the width of each optical luminosity bin and the vertical ones show
the errors in the measured X-ray luminosities.  This plot shows that
the galaxies in our sample follow the established correlation: the
optically-brighter galaxies are also more luminous in the X-rays.  The
existence of this correlation also implies that the detected X-ray
flux from the galaxies in Abell~2634 is not dominated by a few bright
galaxies, but that the optically-fainter galaxies also contribute to
the detected X-ray emission.

The galaxies in Abell~2634 probe the fainter end of the $L_{\rm B}$ --
$L_{\rm X}$ relation as covered by Virgo galaxies.  It should be borne
in mind that there is a bias in the Virgo data which means that the
two data sets in Fig.~6 are not strictly comparable.  At the lower
flux levels, a large number of Virgo galaxies have not been detected
in X-rays, and so this plot preferentially picks out any X-ray-bright
Virgo galaxies.  For the Abell~2634 data, on the other hand, the
co-addition of data from all the galaxies in a complete sample means
that the data points represent a true average flux.  However, it is
clear that the X-ray fluxes from galaxies in these two clusters are
comparable.

The similarity between the X-ray properties of galaxies in these two
clusters is of particular interest because their environments differ
significantly.  The galaxies from the Virgo cluster shown in Fig.~6
lie in a region between 360 kpc and 2 Mpc from the centre of the
cluster.  Recent {\it ROSAT} PSPC observations have shown that the
number density of the hot ICM of this cluster drops from $3 \times
10^{-4}$ to $3 \times 10^{-5} \ {\rm cm^{-3}}$ in this region
(Nulsen \& B\"{o}hringer 1995).  The galaxies from Abell~2634 that
have gone into this plot lie in the inner 0.8 Mpc of Abell~2634, and
in this region the number density of the ICM varies between $1
\times 10^{-3}$ and $2 \times 10^{-4} \ {\rm cm^{-3}}$ (Sakelliou \&
Merrifield 1997). Thus, the galaxies in the current analysis come from
a region in which the intracluster gas density is, on average, an order of
magnitude higher that surrounding the Virgo cluster galaxies.

The location of the galaxies in region (II) of Fig.~6 adds weight to
the tentative conclusion of the previous section that the X-ray
emission from these galaxies can be explained by their X-ray binary
populations, since any significant ISM contribution would place them in
region (I).  Similarly, the low X-ray fluxes of these galaxies leaves
little room for a significant contribution from weak AGN.  If the
X-ray binary populations are comparable to those assumed by Fabbiano
\& Schweizer (1995) in calculating the dividing line in Fig.~6, then
essentially all the X-ray emission from these galaxies can be
attributed to the X-ray binaries.  Thus, any average AGN emission
brighter than a few times $10^{40} \ {\rm erg \ s^{-1}}$ can be
excluded, as such emission would also move the galaxies into region
(I) of the $L_{\rm B}$ -- $L_{\rm X}$ plane.

\subsection{Spiral galaxies}

\begin{figure}
\begin{center}
 \leavevmode
 \epsfxsize 0.9\hsize
 \epsffile{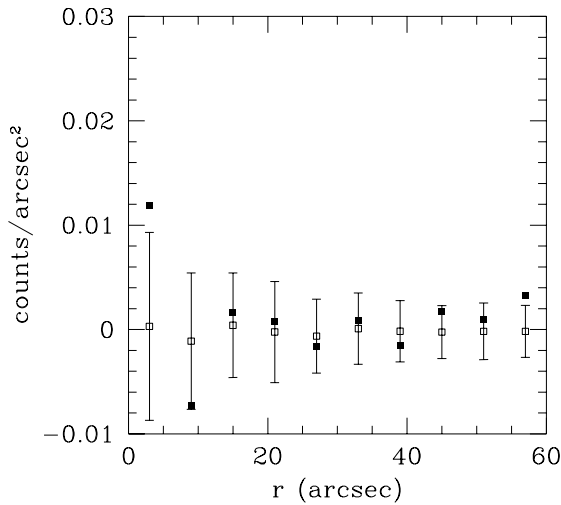}
 \caption{The combined surface brightness profile of the spiral 
galaxies that lie in the field of view of the HRI (filled squares)
normalized to one galaxy. Open squares 
represent the average profile from the simulations.} 
\end{center}
\label{spiral}
\end{figure}

Having discussed the X-ray properties of the early-type galaxies in
Abell~2634 at some length, we now turn briefly to the properties of
the spiral galaxies in the cluster.  Abell~2634 is a reasonably rich
system, and we therefore do not expect to find many spiral galaxies
within it.  Indeed, only 7 of the 62 galaxies whose redshifts place
them at the distance of Abell~2634, and which lie within the field of
the HRI, have been classified as spirals.  The statistics are
correspondingly poor when the X-ray emission around these galaxies is
co-added: the combined profile is shown in Fig.~7, together with the
results from the control simulations (see \S 2.1 for details).  It is
clear from this figure that the spirals have not been detected in this
observation, and a $\chi^2$ fit confirms this impression.

The failure to detect these galaxies is not surprising.  Not only are
there relatively few of them, but their X-ray luminosities are lower
than those of early-type galaxies.  In the {\it Einstein} energy band
(0.2 - 3.5 keV), their luminosities have been found to lie in the
range $\sim 10^{38}$ to $\sim 10^{41} \ {\rm erg \ s^{-1}}$ (Fabbiano
1989).  Modeling this emission using a Raymond-Smith model with a
higher temperature than for the early-type galaxies, as appropriate
for spiral galaxies (Kim et al. 1992a), we find that the expected
count rate for these galaxies is a factor of $\sim 40$ lower than for
the ellipticals in the cluster.  It is therefore unsurprising that we
fail to detect the small number of spiral galaxies present in the
cluster.

\section{Discussion}

In this paper, we have detected the X-ray emission from the normal
elliptical galaxies in Abell~2634.  The limited spatial extent of this
emission coupled with its low luminosity is consistent with it
originating from normal X-ray binaries in the galaxies' stellar
populations.   These galaxies do not seem to have the extended
hot ISM found around galaxies that reside in poorer cluster
environments.  We therefore now discuss whether this difference can be
understood in terms of the physical processes outlined in the
introduction.

Intuitively, the simplest explanation for the absence of an extensive
halo around a cluster galaxy is that it has been removed by ram
pressure stripping as the galaxy travels through the ICM.  A simple
criterion for the efficiency of this process can be obtained by
comparing the gravitational force that holds the gas within the
galaxy to the force due to the ram pressure, which tries to remove it
(Gunn \& Gott 1972).

The gravitational force is given by:
\begin{equation}
F_{\rm GR} \sim G \ \frac{M_{\rm gal} \ M_{\rm gas}}{R_{\rm gal}^{2}}
\end{equation}
where $M_{\rm gal}$ is the total mass of the galaxy, $M_{\rm gas}$ is
the mass of the X-ray emitting gas, and $R_{\rm gal}$ is the radius of
the galaxy's X-ray halo.  For typical values for the masses of the
galaxy and the gas of $10^{12} \ M_{\sun}$ (Forman et al.\ 1985) and
$5 \times 10^{9} \ M_{\sun}$ (e.g.~Canizares et al.\ 1987)
respectively, and a mean value for $R_{\rm gal}$ of $40 \ {\rm kpc}$
(Canizares et al. 1986), which is a representative value for galaxies
of the same optical luminosity as the galaxies in Abell~2634,
equation~(1) implies that $F_{\rm GR} \sim 1 \times 10^{30} \ {\rm
N}$.
 
The force due to ram pressure is described by:
\begin{equation}
F_{\rm RP}=\rho_{\rm ICM} \ v_{\rm gal}^{2} \ \pi R_{\rm gal}^{2} = \mu
\ m_{\rm p} \ n\ v_{\rm gal}^{2} \ \pi R_{\rm gal}^{2}
\end{equation}
where $\rho_{\rm ICM}$ is the density of the ICM, $\mu$ is the mean
molecular weight, $m_{\rm p}$ is the proton mass, $n$ is the number
density of the ICM, and $v_{\rm gal}$ is the galaxy velocity.  From
the velocity dispersion profile of Abell~2634 presented by den Hartog
\& Katgert (1996) we find that the velocity dispersion, $\sigma_{\rm
v}$, in the inner 15 arcmin of this system is $ \approx 710 \ {\rm km
\ s^{-1}}$. Assuming an isotropic velocity field, the characteristic
three-dimensional velocity of each galaxy is hence $v_{\rm gal}=\surd
3 \ \sigma_{\rm v} \approx 1230 \ {\rm km \ s^{-1}}$.  The number
density of the ICM in the same inner region has been derived from
recent {\it ROSAT} PSPC data (Schindler \& Prieto 1997) and this HRI
observation (Sakelliou \& Merrifield 1997), and is found to vary from
$1 \times 10^{-3} \ {\rm cm^{-3}}$ down to $2 \times 10^{-4} \ {\rm
cm^{-3}}$, consistent with previous {\it Einstein} observations (Jones
\& Forman 1984; Eilek et al. 1984).  Inserting these values into
equation~(2), we find $F_{\rm RP} \sim (1 - 10) \times 10^{30} \ {\rm
N}$.

Thus, the ram pressure force exerted on the galaxies in Abell~2634 is
found to be larger than the force of gravity, and so we might expect
ram pressure stripping to be an effective mechanism for removing the
ISM from these galaxies.  In poorer environments, the density of the
ICM is likely to be at least a factor of ten lower, and the velocities
of galaxies will be a factor of $\sim 3$ smaller.  We might therefore
expect $F_{\rm RP}$ to be a factor of $\sim 100$ lower in poor
environments.  Since such a change would make $F_{\rm RP} < F_{\rm
GR}$, it is not surprising that galaxies in poor environments manage
to retain their extensive halos.

The absence of extensive X-ray halos around the galaxies in Abell~2634
implies that ram pressure stripping dominates the processes of
accretion and stellar mass loss which can replenish the ISM.  By
carrying out similar deep X-ray observations of clusters spanning a
wide range of ICM properties, it will be interesting to discover more
precisely what sets of physical conditions can lead to the efficient
ISM stripping that we have witnessed in Abell~2634.

\section*{ACKNOWLEDGEMENTS} 
 
We are indebted to the referee, Alastair Edge, for a very insightful
report on the first incarnation of this paper.  We thank Jason Pinkney
for providing us with the positions and redshifts of the galaxies and
Rob Olling for helpful discussions.  This research has made use of the
NASA/IPAC Extragalactic Database (NED) which is operated by the Jet
Propulsion Laboratory, California Institute of Technology, under
contract with the National Aeronautics and Space Administration. Much
of the analysis was performed using {\sc iraf}, which is distributed
by NOAO, using computing resources provided by STARLINK.  MRM is
supported by a PPARC Advanced Fellowship (B/94/AF/1840).


\begin{thebibliography}{}

  \bibitem[\protect\citename{Awaki et al.} 1994]{awaki} Awaki H.,
Mushotzky R., Tsuru T., Fabian A. C., Fukazawa Y., Loewenstein M.,
Makishima K., Matsumoto H., Matsushita K., Mihara T., Ohashi T.,
Ricker G. R., Serlemitsos R. J., Tsusaka Y., Yamazaki T., 1994, PASJ,
46, L65
  \bibitem[\protect\citename{Bechtold et al.} 1983]{bechtold} Bechtold
J., Forman W., Giacconi R., Jones C., Schwarz J., Tucker W., 
Van Speybroeck L., 1983, ApJ, 265, 26
  \bibitem[\protect\citename{Briel el al.} 1996]{briel} Briel el
al. 1996, $ROSAT$ Handbook
  \bibitem[\protect\citename{Butcher \& Oemler} 1985]{bo} Butcher
H. R., Oemler A. JR., 1985, ApJS, 57, 665
  \bibitem[\protect\citename{Canizares \& Blizzard} 1991]{can91} Canizares
C. R., Blizzard P., 1991, ApJ, 382, 79\
  \bibitem[\protect\citename{Canizares et al.} 1986]{can86} Canizares
C .R., Donahue M. E., Trinchieri G., Stewart G. C., McGlynn T. A.,
1986, ApJ, 304, 312 
  \bibitem[\protect\citename{Canizares et al.} 1987]{can87} Canizares
C. R., Fabbiano G., Trinchieri G., 1987, ApJ, 312, 503
  \bibitem[\protect\citename{den Hartog \& Katgert} 1986]{} den Hartog
R., Katgert P., 1996, MNRAS, 279, 349
  \bibitem[\protect\citename{Eilek et al.} 1984]{eilek} Eilek J. A.,
Burns J. O., O'Dea C. P., Owen F. N., 1984, ApJ, 278, 37
  \bibitem[\protect\citename{Fabbiano } 1989] {} Fabbiano G., 1989,
ARA\&A, 27, 87
  \bibitem[\protect\citename{Fabbiano \& Schweizer} 1995]{} Fabbiano
G., Schweizer F., 1995, ApJ, 447, 572
  \bibitem[\protect\citename{Fabbiano et al.} 1992]{fab92} Fabbiano
G., Kim D.-W., Trinchieri G., 1992, ApJS, 80, 531 
  \bibitem[\protect\citename{Forman et al.} 1985]{forman85} Forman W.,
Jones C., Tucker W., 1985, ApJ, 293, 102
  \bibitem[\protect\citename{Forman et al.} 1979]{forman79} Forman W.,
Schwarz J., Jones C., Liller W., Fabian A. C., 1979, ApJ, 234, L27
  \bibitem[\protect\citename{Grebenev et al.} 1995]{grebenev} Grebenev
S. A., Forman W., Jones C., Murray S., 1995, ApJ, 607
  \bibitem[\protect\citename{Gunn \& Gott} 1972]{gunngott} Gunn J. E.,
Gott J. R., 1972, ApJ, 176, 1
  \bibitem[\protect\citename{Jones \& Forman} 1984]{} Jones C., Forman
W., 1984, ApJ, 276, 38 
  \bibitem[\protect\citename{Kim et al.} 1992a]{kim92a} Kim D.-W.,
Fabbiano G., Trinchieri G., 1992a, ApJS, 80, 645 
  \bibitem[\protect\citename{Kim et al.} 1992b]{kim92b} Kim D.-W.,
Fabbiano G., Trinchieri G., 1992b, ApJ, 393, 134
  \bibitem[\protect\citename{Kormendy \& Richstone} 1995]{} Kormendy
J., Richstone D., 1995, ARA\&A, 33, 581
  \bibitem[\protect\citename{Kormendy et al.} 1996a]{} Kormendy J.,
Bender R., Ajhar E. A., Dressler A., Faber S. M., Gebhardt K.,
Grillmair C., Lauer T. R., Richstone D., Tremaine S., 1996a ApJL, 473,
91 
  \bibitem[\protect\citename{Kormendy et al.} 1996b]{} Kormendy J.,
Bender R., Ajhar E. A., Dressler A., Faber S. M., Gebhardt K.,
Grillmair C., Lauer T. R., Richstone D., Tremaine S., 1996b ApJL, 459,
57 
  \bibitem[\protect\citename{Mahdavi et al. }1996] {} Mahdavi A.,
Geller M. J., Fabricant D. G., Kurtz M. J., 1996, AJ, 111, 64
  \bibitem[\protect\citename{Matsushita et al.} 1994]{mmm} Matsushita
K., Makishima K., Awaki H., Canizares C. R., Fabian A. C., Fukazawa
Y., Loewenstein M., Matsumoto H., Mihara T., Mushotzky R. F., Ohashi
T., Ricker G. R., Serlemitsos P. J., Tsuru T., Tsusaka Y., Yamazaki
T., 1994, ApJ, 436, L41
  \bibitem[\protect\citename{Merritt} 1983]{merritt83} Merritt D.,
1983, ApJ, 264, 24
  \bibitem[\protect\citename{Merritt} 1984]{merritt84} Merritt D.,
1984, ApJ, 276, 26
  \bibitem[\protect\citename{Nulsen \& B\"{o}hringer} 1995] {} Nulsen
P. E. J., B\"{o}hringer H., 1995, MNRAS, 274, 1093
  \bibitem[\protect\citename{Oemler} 1974]{oe} Oemler A., 1974, ApJ,
194, 1
 \bibitem[\protect\citename{Pinkney } 1995]{pinkney95} Pinkney J.,
1995, PhD Thesis, New Mexico State University
  \bibitem[\protect\citename{Pinkney et al.} 1993]{pinkney93} Pinkney
J., Rhee G., Burns J. O., Hill J. M., Batuski D., Hintzen P., 
1993, ApJ, 416, 36
  \bibitem[\protect\citename{Rangarajan et al.} 1995]{rangarajan}
Rangarajan F. V. N., White D. A., Ebeling H., Fabian A. C., 1995, 
MNRAS, 277, 1047
  \bibitem[\protect\citename{Raymond \& Smith} 1977]{raysmith} Raymond
J. C., Smith B. W., 1977, ApJS, 35, 419 
  \bibitem[\protect\citename{Richstone} 1975]{richstone}  Richstone
D. O., 1975, ApJ, 200, 535
  \bibitem[\protect\citename{Sakelliou \& Merrifield} 1997]{}
Sakelliou I., Merrifield M. R., 1997, in preparation
  \bibitem[\protect\citename{Sakelliou et al.} 1996]{sakelliou}
Sakelliou I., Merrifield M. R., McHardy I. M., 1996, MNRAS, 283, 673  
  \bibitem[\protect\citename{Schindler \& Prieto} 1997] Schindler S.,
Prieto M. A., 1997, A\&A, accepted for publication
  \bibitem[\protect\citename{Scodeggio et al.} 1995]{scodeggio}
Scodeggio M., Solanes J. M., Giovanelli R., Haynes M., 1995, 444, 41
  \bibitem[\protect\citename{Soltan \& Fabricant} 1990]{soltan} Soltan
A., Fabricant D. G., 1990, ApJ, 364, 433
  \bibitem[\protect\citename{Stark et al.} 1992]{stark} Stark A. A.,
Gammie C. F., Wilson R. W., Bally J., Linke R. A., Heiles 
C., Hurwitz M., 1992, ApJS, 79, 77
  \bibitem[\protect\citename{Trinchieri \& Fabbiano} 1985]{trinfab85}
Trinchieri G., Fabbiano G., 1985, ApJ, 296, 447
  \bibitem[\protect\citename{van der Marel et al. } 1997]{} van der
Marel R. P., de Zeeuw P. T., Hans-Walter Rix, Quinlan G. D., 1997,
Nat, 385, 610 
  \bibitem[\protect\citename{Vikhlinin et al.} 1994]{vikhlinin}
Vikhlinin A., Forman W., Jones C., 1994, ApJ, 435, 162
  \bibitem[\protect\citename{White et al.} 1991]{white91} White D. A.,
Fabian A. C., Forman W., Jones C., Stern C., 1991, ApJ, 375, 35
  \bibitem[\protect\citename{Wrobel} 1991]{} Wrobel J. M., 1991, AJ,
101, 127

\end{thebibliography}
\end{document}